\documentclass{article}

\usepackage{amsmath}
\usepackage{amsfonts}
\usepackage{graphicx}
\usepackage{mathtools}
\usepackage{verbatim}
\usepackage[hidelinks]{hyperref}
\usepackage{comment}
\usepackage{xcolor}

\newtheorem{theorem}{Theorem}[section]

\numberwithin{equation}{section}

\begin{document}

\title{Evolution of the propagator matrix method and its implementation in seismology}

\author{Thomas Meehan\footnote{Department of Earth Sciences, Memorial University of Newfoundland, St. John's
, NL A1C 5S7 \texttt{tbm247@mun.ca}}}

\maketitle
\begin{abstract}
In this paper, we review development of an algorithm that is referred to in seismology as the Haskell matrix method, the Thomson-Haskell matrix method, or the propagator matrix method.
The roots of this algorithm and main developments are examined to offer a better understanding of its essential features.
The underlying theory is highlighted by removing specific expressions and manipulations that often shroud the common method involved. 
Also, I discuss implementations of the algorithm in Python, with a reference to source code.
These implementations calculate dispersion curves for guided waves.
\end{abstract}

\section{Introduction}

The Haskell layer matrix method is known to researchers in seismology from Haskell's~\cite{Ha} paper received by the Bulletin of the Seismological Society of America on June 5, 1951. Haskell credits the work of Thomson~\cite{Ta} in his formulation of the method. The methodology has roots in vibration analysis, is related to the product integral, and finds wider use in seismology as computational machinery becomes available. 

The literature that has grown through the use of this method has become extensive and a complete review would be difficult. A presentation of the underlying theory, however, is straightforward. The theory seen in a general setting offers reasonable clarity, which can often be clouded by the specifics of a particular formulation. 

Following this method, from its inception to the present time, a growing applicability can be observed due to the increasing use of modern computers. The computer programming environment has also changed throughout the years, creating opportunity for collaborative programming and the adaptation of codes to improving technologies, which is a benefit to researchers. Coding is emerging from an \textit{obscure art} to a \textit{lingua franca} of technology and science---it is not surprising if one recognizes this art as a dialect of mathematics. 

\section{Matrix method}

A first description of the matrix method is given without reference to any particular model of a physical system. Following the initial description, a selection of particular applications are discussed, emphasizing the method's early roots with reference to pertinent literature. A review of the literature builds a solid foundation on which the matrix method for dispersion relations of guided waves in multilayer media can be easily understood.

\subsection{General formulation}
\subsubsection{Introduction}
Herein, I describe a five-step procedure in the formulation of a transfer matrix $\mathbf{P}$ and a propagator matrix $\mathbf{T}$. Herein, bold entities refer to matrix quantities. This is not an exhaustive procedure and is provided to highlight the main mathematical relationships of the matrix method.
\subsubsection{Constructing the propagator}
\begin{enumerate}

\item A mathematical model of physical systems in one spatial dimension, with the following characteristics, can be created by dividing the system into $n$ sections. This division naturally creates $n+1$ boundaries. Each section is described by equations of equivalent form, with parameters $p_1^k,p_2^k,\ldots,p_{n_p}^k$  and end conditions $C_1^i,C_2^i,\ldots,C_{n_c}^i$, where the values of the parameters and end conditions are dependent on the $k^{th}$ section and $i^{th}$ boundary respectively. The subscripts $n_p$ and $n_c$ represent the number of parameters and end conditions, respectively. 

\item The equations of any one section are then solved for the end conditions $C_1^i,C_2^i,\ldots,C_{n_c}^i$ giving a set of equations,

\begin{align*}
 C_1^i& = f(p_1^k,p_2^k,  \ldots,p_{n_p}^k, C_1^{i+1},C_2^{i+1}\ldots,C_{n_c}^{i+1}) \\
 C_2^i& = f(p_1^k,p_2^k,  \ldots,p_{n_p}^k, C_1^{i+1},C_2^{i+1}\ldots,C_{n_c}^{i+1}) \\
 & \vdotswithin{=} \\
 C_{n_c}^i& = f(p_1^k,p_2^k,  \ldots,p_{n_p}^k, C_1^{i+1},C_2^{i+1}\ldots,C_{n_c}^{i+1}).
\end{align*}

\item This system of equations can then be represented in a matrix formulation. The left hand side is made into a $n_c \times 1$ vector, which is typically called the state vector. The right hand side can be decomposed into a coefficient matrix of the parameters $p_1^k,p_2^k,\ldots,p_{n_p}^k$ and a vector of the other set of end conditions $C_1^{i+1},C_2^{i+1},\ldots,C_{n_c}^{i+1}$. This vector has the same shape as the vector on the left hand side. The coefficient matrix, which is referred to as a transfer matrix, relates the state vectors of the two end conditions of a section and is a square matrix of $n_c \times n_c$. The matrix formulation is given by,

\begin{equation*}
\left[ \begin{array}{c} C_1^i \\ \vdots \\ C_{n_c}^i \end{array} \right] = \begin{bmatrix} P_{11} & \cdots & P_{1n_c} \\ \vdots & \ddots & \vdots \\ P_{{n_c}1} & \cdots & P_{{n_c}{n_c}} \end{bmatrix} \left[ \begin{array}{c}C_1^{i+1} \\ \vdots \\ C_{n_c}^{i+1} \end{array} \right],
\end{equation*}
where the elements of the coefficient matrix are arbitrary functions of the parameters~$p_1^k,p_2^k,\ldots,p_{n_p}^k$ depending on the mathematical description. The above formulation can be represented as,

\begin{equation*}
\mathbf{C}_i = \mathbf{P}_k \times \mathbf{C}_{i+1},
\end{equation*}
where the superscripts can now be moved to subscripts.  

\item Taking all $n$ sections into account we have a system of equations in direct matrix notation,
\begin{align*}
\mathbf{C}_1 &= \mathbf{P}_1 \times \mathbf{C}_2\\
\mathbf{C}_2 &= \mathbf{P}_2 \times \mathbf{C}_3\\
& \vdotswithin{=} \\
\mathbf{C}_{n-1} &= \mathbf{P}_n \times \mathbf{C}_n.
\end{align*}

\item Thus, we can relate conditions $\mathbf{C}_1$ to conditions~$\mathbf{C}_n$ through a multiplication of matrices $\mathbf{P}_k$,

\begin{equation*}
\mathbf{C}_1 = \mathbf{P}_1 \times \mathbf{P}_2 \times \ldots \times \mathbf{P}_n \times \mathbf{C}_n.
\end{equation*}
Letting,
\begin{equation*}
\mathbf{T} = \mathbf{P}_1 \times \mathbf{P}_2 \times \ldots \times \mathbf{P}_n,
\end{equation*}
where $\mathbf{T}$ is now a lumped parameter system consisting of the parameters each section, we obtain
\begin{equation} \label{eq:1}
\mathbf{C}_1 = \mathbf{T} \times \mathbf{C}_n.
\end{equation}
This is a simplified form; in applications many alterations are possible leading to complicated expressions.
\end{enumerate}
\subsubsection{Final expressions}
The details of these five steps are dictated by the specific expressions of the mathematical model, the information sought, and the possible manipulations to achieve this form. The matrix $\mathbf{P}$ ``transfers" information between the end conditions of a section. The matrix $\mathbf{T}$ ``propagates" this information from one section to another. 

\subsection{Application}
\subsubsection{First seed}
The skeletal structure of the matrix methods  can be seen in the applications of vibrational analysis in the works of Myklestad~\cite{M} and Timoshenko~\cite{TS}. These texts do not have any explicit matrix methods, however, their brief description of a tabular method to calculate the natural frequencies of vibrational systems is likely the seed of later matrix methods. Both authors attribute the origination of these techniques to Heinrich Holzer. 
\subsubsection{Fully formed method}
The first instance of the matrix method in available literature is in the textbook of Thomson~\cite{Tb}. A transfer matrix is introduced for a spring-mass system in a section of the text entitled ``a Holzer-Type problem". Shortly thereafter, Thomson~\cite{Tc} published an article, which uses a transfer matrix for a model of a column with varying width and is able to calculate the critical load. A transfer matrix for layered media is also derived in a second article by Thomson~\cite{Ta}. These instances demonstrate the use of the fully-formed method. 
\subsubsection{Well-developed theory}
The matrix method is further developed by Pestel and Leckie~\cite{P}, where Thomson is mentioned in the preface as a source of encouragement. The authors find many applications to these methods in the context of engineering. The last chapter provides a catalogue of transfer matrices for straight, curved and twisted beams, and rotating disks. The seventh chapter introduces the delta matrix method, which is a variant of the above generalized method; the delta matrix method is used in seismology and the computer codes I developed in this investigation.
\subsection{Dispersion relations}
\subsubsection{Introduction to seismology}
The matrix method allows one to extract particular information from the mathematical model. Thomson~\cite{Ta} assumes overall boundary conditions for a liquid medium, deriving expressions for reflection and transmission coefficients. The state vectors of end conditions consist of two particle velocity terms and two stress terms. The transfer matrix contains the parameters of frequency, isotropic elasticity, and the thickness of the medium layers. This is the first instance of the matrix method being applied to a layered earth model.
\subsubsection{Multilayer dispersion}
Haskell~\cite{Ha}, following Thomson~\cite{Ta},  produces an equivalent form\footnote{Haskell made a correction in Thomson's formulation and uses a different form for the direction angles of the wave normal.} of the transfer matrix and utilizes it to produce dispersion relations for quasi-Rayleigh waves while also providing a derivation of a transfer matrix for Love waves. Haskell~\cite{Ha} indicates the dispersion relation for multiple layers is treated in Sezawa~\cite{Se}, but makes note of the increased efficiency of the matrix notation.

In a model of the earth as flat parallel layers, Haskell~\cite{Ha} uses the boundary conditions of no stresses at the free surface and no sources at infinity to represent the matrix product as a ratio of terms, relating velocity $v$, and wave number $k$. The matrix product is the product of transfer matrices, as generalized above, in equation \ref{eq:1}. Each Earth-model layer will have elasticity and thickness parameters, as mentioned in section 2.3.1, specific to each layer. An iterative scheme is used to find the velocity and wave number pair $(v,k)$, which is known as the dispersion relation.

This method allows the extension to any number of layers, where the only limit is the computational resources available. In the case of a single layer over a halfspace, the method reduces to previously found expressions. Stoneley~\cite{st} and Buchen and Ben-Hador~\cite{B} discuss limiting cases. In the long-wavelength limit, the transfer matrix  approaches identity and the expression reduces to the Rayleigh equation for a halfspace. The high-frequency limit allows the matrix product to be factored into multiple terms representing a Rayleigh wave in the uppermost layer and Stoneley waves at the interfaces below.

\subsubsection{Model variations}
Haskell~\cite{Hb,Hc} examines radiation patterns from a fault in a homogeneous medium, then develops the matrix method dealing with a point source between layers. A growing number of researchers in seismology have begun to utilize the methodology for Earth-layer models. Harkrider~\cite{HR} uses the method to model buried sources. Anderson~\cite{A} investigates transversely isotropic formulations and is later joined by Harkrider~\cite{HAN}. Woodhouse~\cite{Wo} looks into layers with varying thickness. In all cases researchers develop a mathematical description in the form indicated above in equation (\ref{eq:1}). This development can become convoluted and could shroud the simplicity of the underlying method.

\section{Product Integration}

The transfer matrix $\mathbf{P}$ and its product $\mathbf{T}$, from equation (\ref{eq:1}), is related to the theory of product integrals in the work of Gilbert and Backus \cite{GB}. They describe the matrix method as an ``approximant" to the product integral. This insight connects the algorithm to a larger body of mathematics and $\mathbf{T}$ is referred to as a propagator matrix.

Dollard and Friedman \cite{PI} present a modern theory of product integration. Herein, some elementary results from this theory are reproduced to demonstrate how it is connected to the matrix method.

\subsection{Approximation to product integration}
\subsubsection{Riemann analogue for product integration}
In a comparison to the above generalized form of the matrix method, let us consider the system of differential equations,   
\begin{align*}
 y'_1(s)& = a_{11}(s)y_1(s) + a_{12}(s)y_2(s) + \ldots + a_{1n}y_n(s) \\
 & \vdotswithin{=}  \\
 y'_{n}(s)& = a_{n1}(s)y_1(s) + a_{n2}(s)y_2(s) + \ldots + a_{nn}y_n(s).
\end{align*}
The coefficients $a_{ij}(s)$ correspond to functions of the parameters $p_i$ while the functions of $y_j(s)$ correspond to the end conditions $C_i$ from step (3). The functions representing the elements of the transfer matrix and state vectors, respectively, are both dependent on $s$, which corresponds to the single spatial dimension of the physical system. Expressing this system in matrix form we have,
\begin{equation} \label{eq:2}
\mathbf{Y'}(s)=\mathbf{A}(s)\mathbf{Y}(s).
\end{equation}

A solution $\mathbf{Y}(s_n)$ is obtainable, if an initial value $\mathbf{Y}(s_0)$ is known. The approximate solution is provided in an analogous manner to the Riemann integral for ordinary integrals. A partition $P = \{s_o,s_1,\ldots,s_n\}$ is created for the interval $[s_0,s_n]$. A constant value is chosen from each sub-interval $\mathbf{A}(s_i,s_{i-1})$, yielding $\mathbf{A}(s_i)$, which turns $\mathbf{A}(s)$ into a step function. The approximate solution at each step is given by solving equation (\ref{eq:2}) to obtain

\begin{equation*}
\mathbf{Y}(s_{i+1}) \cong e^{\mathbf{A}(s_i)(s_i-s_{i-1})}\mathbf{Y}(s_i).
\end{equation*}

Starting at the initial value $\mathbf{Y}(s_0)$ and using a similar recursion scheme as given in the early formulation, a solution for $\mathbf{Y}(s_n)$ is approximated as

\begin{equation*}
\mathbf{Y}(s_n) \cong e^{\mathbf{A}(s_n)\Delta s_n} \ldots e^{\mathbf{A}(s_1)\Delta s_1}Y(s_0) \cong \prod_{k=1}^n e^{\mathbf{A}(s_k)\Delta s_k}Y(s_0).
\end{equation*}
This expression can be related to the matrix method by associating the matrix exponential with the transfer matrix and their product with the matrix $\mathbf{T}$. Taking a constant value for each sub-interval corresponds to dividing the earth layer model described in section 2.3.2 into layers of constant properties.
\subsubsection{Taking the limit to obtain the product integral}
The product of matrix exponentials is denoted as,
\begin{equation}
{\prod} _P(\mathbf{A}) \equiv \prod_{k=1}^n e^{\mathbf{A}(s_k)\Delta s_k}.
\end{equation}
  The product integral is defined as
\begin{equation}
\mathbf {\lim_{\mu(P)\to 0}{\prod} _P(A)},
\end{equation}
where $\mathbf \mu(P)$ refer to the length of the longest sub-interval.
\subsubsection{Product integral as a propagator}
One can see that the product integral is a propagator from the above derivation or by examining its multiplicative property, which is analogous to the additive property of the ordinary integral. 

\begin{theorem}[Multiplicative Property]
Let $\mathbf{A:[a,b]\to\,}  \mathbb{C}_{n \times n}$ be continuous, and let $\mathbf{x,y,z \in [a,b]}.$ Then
\begin{equation*}
\prod_z^x e^{\mathbf{A}(s)ds} = \prod_y^x e^{\mathbf{A}(s)ds}\prod_z^y e^{\mathbf{A}(s)ds}.
\end{equation*}
\end{theorem}

The proof of Theorem 3.1 is provided in Dollard and Friedman \cite{PI}. Therein, the authors describe a propagator as ``an operator that, acting on the value of a solution at one point, produces the value at another point". The propagator generalizes the two previously described matrices $\mathbf{P}$ and $\mathbf{T}$. 

\subsection{Product integral in seismology}
\subsubsection{New mathematical tool}
The product integral offers an alternative way to formulate the expression required for the transfer matrix. The expression, often obtained through the calculation of a matrix inverse, can now be converted to the evaluation of the matrix exponential. This evaluation is essentially an eigenvalue problem and as stated by Udias~\cite{U}, ``there are standard fast methods for calculating the eigenvalues and eigenvectors of a matrix". The product integral provides alternative formulations for the matrix method and opens new theory to the analyst. 
\subsubsection{Specific use in seismology}
A first instance of the propagator matrix formulation is Hudson~\cite{HU} in a investigation of seismic sources. The propagator formulation becomes extensively utilized. However, as the mathematical theory of propagator matrices is not well developed at the time, the notation used to describe this theory could vary considerably. Of particular note is the work of Kennett~\cite{K}, which provides a comprehensive treatment of the subject for layered media. Kennett refers to the text as ``a unified account of seismic waves in stratified media", and this account is largely based on propagator matrices.  

\section{Computer Codes}
Some of the first computing machinery available had algorithms implemented, utilizing the matrix method in seismology. This work continues to the present day with different computer platforms and languages. This demonstrates the utility of the methods, but also allows one to view the effect of evolving computation equipment and algorithm implementation. The difficulties and benefits associated in working with this fast evolving medium are evident in the research papers produced. 

The computer codes created could be viewed as the tip of a fast moving wave, starting from a Haskell point source. Using the work of multiple formulations presented by researchers in this domain, Python code is developed to calculate dispersion curves for guided waves. This particular work was conducted to assist The Geomechanics Project at Memorial University of Newfoundland in an investigation of assumptions made in the utilization of the Backus average. The code resulting from this work are presented on GitHub at \href{https://github.com/tbmcoding/dispersion}{\color{gray!20!blue!90}https://github.com/tbmcoding/dispersion}. 

\subsection{Evolution of the algorithm}
\subsubsection{Early implementations}
The dispersion algorithm using the Haskell matrix method was first put to use on the Colombia University IBM 650 computing machine by researchers Dorman et al.~\cite{DEO}. They describe computing 256 different quasi-Rayleigh wave phase velocities for 11 different models with up to 50 layers. This is an early instance of, what is commonly referred to in seismology as, inversion by trial and error. This forward modelling technique is described in Dorman et al.~\cite{DEO} in the following quote.
\begin{quote}
On the other hand, the surface-wave technique used here is indirect, i.e., dispersion is computed for assumed models in an attempt to fit observed dispersion.
\end{quote}

A second implementation by Press et al.~\cite{Sea}, first working on a Bendix G-15D, required an overnight run for a typical 20 layer model. They subsequently implemented code on an IBM 704, and finally, an IBM 7090, as well as offering a mail-order service for computation. The mail-order service is an indication of the type of collaboration available to researchers at this stage of the algorithm's progression.

\subsubsection{Computational difficulties}
In the early implementations the limitations due to machine overflow and precision that existed in the Haskell matrix method were recognized. Researchers develop new variants of the Haskell matrix method that solve computational difficulties. The computational problem is described in detail by Dunkin~\cite{D}. The efficiency of the developed algorithms is taken to be an important aspect, as seen in a comparison of these variations by Buchen and Ben-Hador~\cite{B}. 
\subsubsection{Recent implementations}

New implementations have been numerous, as seen by Chapman's~\cite{CHAP} aptly named paper: \textit{Yet another elastic layer plane-wave, layer-matrix algorithm}. His paper outlines code in the high-level language Matlab and is the first to include actual code. To paraphrase Chapman, the consideration of program efficiency has become less of a concern with the increase in the speed of computational machinery. However, if processing a large number of models was of concern, efficiency could again become important.

Two of the most recent implementations are Ke et al.~\cite{ke} and Ikeda and Matsuoka \cite{Ike}. The first paper derives another method to eliminate the precision problem and a new accelerated root searching scheme. The second paper derives the reduced delta matrix formulation reviewed in Buchan and Ben-Hador but for transversely isotropic media. This formulation is utilized in the coding provided on GitHub.

\subsection{Developed Codes}
\subsubsection{Language considerations}
Using a collection of the previously developed formulations, the codes provided on the GitHub repository can compute dispersion curves for guided waves, specifically quasi-Rayleigh and Love waves, for media with isotropic and transversely isotropic symmetry. The Python language was chosen due to its flexibility and use in other recent projects in seismology.

Python offers a fully object-oriented environment and the added advantage of being open source. Object orientation allows rapid prototyping through modularization, leading to efficient code testing and bug removal. Additional support can be added for program efficiency through vectorization, modular C implementation, or parallel programming \cite{DP}. The question of open source could be more of a philosophical or, perhaps, economic choice but does lend itself to the academic spirit of collaboration.
\subsubsection{Isotropic code}
For the isotropic quasi-Rayleigh wave code, the Thomson-Haskell formulation presented in the review of Buchen and Ben-Hador~\cite{B}, appendix A2, was followed. A symbolic comparison between the results of the Thomson-Haskell formulation, for the case of a single layer over a halfspace, is made to the determinant equation that is derived by Dalton et al.~\cite{DS}. The comparison was found to be equivalent to a multiplicative factor confirming the accuracy of the expressions. The unified notation of the multiple formulations that are reviewed in Buchen and Ben-Hador~\cite{B} enabled an easy extension to the delta matrix representation, which solved the precision problems discussed in section 4.1.2. The unified notation also allows a slight modification of the existing code to handle Love-wave dispersion.
\subsubsection{Transversely isotropic code}
For the transversely isotropic code, the formulations of Ikeda and Matsuoka~\cite{Ike} are used in conjunction with the matrix representation of Buchen and Ben-Hador~\cite{B}. This formulation allows a similar algorithm implementation that proves to be efficient for later vectorization. Given the transversely isotropic equivalent parameters for a isotropic medium, the expressions reduce to the isotropic case. A comparison to previously developed isotropic code is then possible. For Love-wave transversely isotropic codes, a pseudo-rigidity and pseudo-thickness is substituted into the Buchen and Ben-Hador \cite{B} formulations according to the derived expressions from Anderson \cite{An}.
\subsubsection{Implementation and efficiency}
Initially, I utilize the Sympy module to develop code with matrix function capabilities. This provides straightforward algorithm implementation and highly readable code. The arbitrary precision feature of the Python language adds the additional benefit of eliminating concern for machine overflow. The efficiency is moderate due to multiple layers consuming significant computation time.

In a second implementation, I utilize the Numpy module with vectorization and broadcasting to reduce computation time. Three-dimensional arrays are created in a frequency $\times$ velocity $\times$ layer space and consolidation of all repeated calculations. The computation time is significantly reduced allowing the fast exploration of model changes.

\section*{Acknowledgments}
I wish to acknowledge email communication with Tatsunori Ikeda, and the editorial work of David Dalton, Elena Patarini, Michael Slawinski and Theodore Stanoev. This research was performed in the context of The Geomechanics Project supported by Husky Energy.
Also, this research was partially supported by the Natural Sciences and Engineering Research Council of Canada, grant 238416-2013.
   
\bibliography{Evo.bib}
\bibliographystyle{unsrt}

\end{document}